# FPGA-Based Material Testing Machine Controller

Arev Hambardzumyan, Rafayel Ghasabyan, Vahagn Tamazyan

TACTUN INC, USA

Emails: arevh@tactun.com, rafayelg@tactun.com, vahagnt@tactun.com

*Abstract*— **In the realm of contemporary materials testing, the demand for scalability, adaptability, parallelism, and speed has surged due to the proliferation of diverse materials and testing standards. Traditional controller-based systems often fall short in meeting these requirements, resulting in adaptability and processing speed limitations. Conversely, FPGA-based controllers present a multifaceted, high-performance solution. Key advantages of FPGA-based controllers in materials testing encompass reconfiguration capabilities for cost-effective adaptation to evolving materials and standards. FPGAs also enable the integration of parallel control and data acquisition circuits, vital for multichannel test equipment demanding simultaneous, independent operation of multiple control channels.**

*Keywords— FPGA, IoT, smart machinery, Industry 4.0, Industry 5.0, automation, control, multi-axis, PID, data logging, data acquisition, motion control, material testing.*

## I. INTRODUCTION

Destructive material testing equipment serves a broad range of industries, encompassing automotive, medical, construction, consumer goods, and scientific institutions, among others. A wide range of mechanical testing equipment is accessible, offering versatility in evaluating parameters like compression, adhesion, flexure, fatigue, shock, elasticity, vibration, tensile strength, and shear, ensuring rigorous quality assessment of the tested materials.

Modern materials testing requires scalability, adaptability, parallelism, and speed. With the advance in materials sciences, the variety of materials rapidly increases every day. Such an increase in the variety of materials drives a similar increase in the number of standards for testing these new materials. Because the new materials are more advanced, such materials may have multiple different properties for testing, multi-vector test scenarios, more precise and complex calculations, and/or higher rate of changes in test scenarios.

More importantly, some materials require very fast feedback during testing. The testing standard for materials may require adjustments to the testing of the materials based on the current condition (e.g., active feedback loop). Such materials test requires speed and/or parallelism in data processing, calculations, and output signal generation.

Additionally, the hardware of the materials test equipment is costly; thus, it is not cost-effective to change/replace the hardware of the materials test equipment for each new standard and/or material. Accordingly, flexible architecture is required to adapt to the same materials test equipment for the variety of materials and/or standards.

## II. PROPOSED SOLUTION

Materials test equipment usually is controlled by generic microcontroller-based systems, but as described above, such controllers may lack adaptability, parallelism, and speed. In contrast, patented [1] techniques proposed herein include FPGA-based (Field Programmable Gate Array) controllers that do not have the limitations of such controllers and improve the performance of the test equipment while providing more functions for industrial and research applications.

FPGAs have the adaptability and flexibility to combine the strengths of processor and DSP (Digital Signal Processing) and may be used to target a wide range of testing applications.

The main advantages of the FPGAs in materials testing applications are:

- Reconfiguration: using techniques described herein, FPGA may be reconfigured at any time, providing cost-efficient adaptability by updating for materials to support various research and industrial applications in the future.

- Parallelism: using techniques described herein, parallel control and data acquisition circuits may be incorporated into an FPGA for running parallel functions simultaneously. This is especially important for multichannel test equipment, where several control channels should work in parallel, independently, and simultaneously.

- High speed: using techniques described herein, circuits may be implemented on FPGA hardware to run at a fast clock rate. Combined with the parallel execution of functions within the FPGA, the speed and data throughput requirement for multichannel control and data acquisition may be easily met.

In an embodiment, main materials test control and data acquisition functions are performed by an FPGA-based controller. The use of the FPGA-based controllers provides the highest possible accuracy, control speed, and reliability.

FPGA-based controller functions include one or more:

- Data acquisitions (implementation of drivers for the analog-to-digital converters (ADC) or analog front-ends (AFE) for high-speed data digitization),
- Control/output signal (waveform) generation,
- Feedback control loop implementation,
- Actuator control signal generation (implementation of driver for the digital-to-analog converters (DAC) for high-speed control of actuators, implementation of pulse-width modulation (PWM) signals for actuator control),
- Digital signal monitoring, and/or
- Auxiliary input/output control.

Other functions of controllers for materials testing equipment may be implemented by a Central Processing Unit (CPU), including communication with the FPGA and network-coupled computing devices PC/Tablet/Cloud, data logging, high-level control of the testing process, etc.

### III. SOLUTION OVERVIEW

#### A. System Overview

Material Test Systems (MTS) differ with the types of tests they can perform, number of actuators and sensors (test devices) included in the machine, and number of machines included in the system, which are controlled from the same controller.

Actuator Device may be any type of motor that performs angular or linear motion. Non-limiting examples of actuators devices for destructive and non-destructive testing include servo-hydraulic, electro-mechanical, electromagnetic, ultrasound, and laser-based actuators. Actuator receives output signals from Test Controller to produce a movement that affects a sample material according to test requirements.

Sensor device may be a sensor that collects the measurements of the effects of a materials test or the measurements of the sources that produce the effects. For example, a sensor device may be a load cell sensor measuring the load force on a sample material. As another example, sensor device may be an encoder coupled to Actuator Device that measures the displacement.

For example, multi-actuator test machines may be used for multi-axis materials testing, in which each actuator is affecting different motions for the materials testing on the same sample material. Such a configuration is particularly useful for large-scale materials testing (e.g., wind turbine blade specimen).

FIG. 1 is a block diagram that depicts Material Test System, which includes Materials Test Management System, i.e. controller based on FPGA and Materials Test Machine. Materials Test Machine is a hardware machine that performs destructive and/or non-destructive materials tests on sample material.

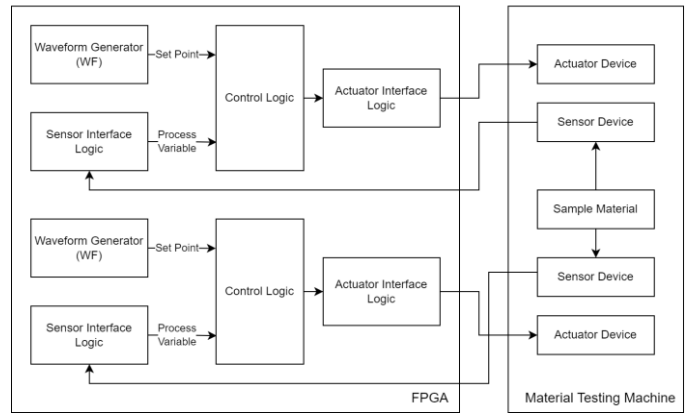

Fig. 1. Material test system diagram

Materials Test Management System manages Materials Test Machine, including deployment, provisioning, and configuring Materials Test Machine and its parts. Materials Test Management System may configure the tests to be performed on sample materials and collect/analyze the results of the tests. Materials Test Management System interfaces with Materials Test Machine through Test Controller. The Test Controller manages the execution of tests.

This particular system depicted in Fig. 1 includes two actuators, and thus the Materials Test Management System includes two parallel paths for the control of two actuators independently.

The control Logic block depicted on the diagram can implement various control algorithms, such as Proportional, Integral, Derivative (PID) close-loop control, Feed-Forward control, Fuzzy logic, etc. Additional algorithms which can be configured are adaptive amplitude and mean control algorithms.

The Waveform Generator (WF) modules generate high accuracy 64-bit profiles for the actuator control. The module can generate various profiles covering needs of different tests, such as static tests, dynamic/fatigue tests, vibration tests, etc. Types of profiles include but are not limited to ramp profiles, sine-wave, square-wave, triangular, random-sine, tapered sine-wave, sweep sine-wave, etc.

#### B. Functional Overview

FIG. 2 is a flowchart that depicts a process for performing materials testing with FPGA-based MTS.

Test Controller receives configuration data for configuring Test Controller, including FPGA, and Materials Test Machine, including its test devices. CPU of Test Controller processes configuration data, including machine config data and/or test config data, to determine configuration setting data for FPGA.

The configuration settings may include parameters for configuring communication channels between Test Controller and Materials Test Machine (e.g., which channel endpoints are connected to which materials testing device), configuring interface components (e.g., ADCs and DACs), configuring control signals (e.g., for generating waveform causing a corresponding movement of Actuator Device), etc.

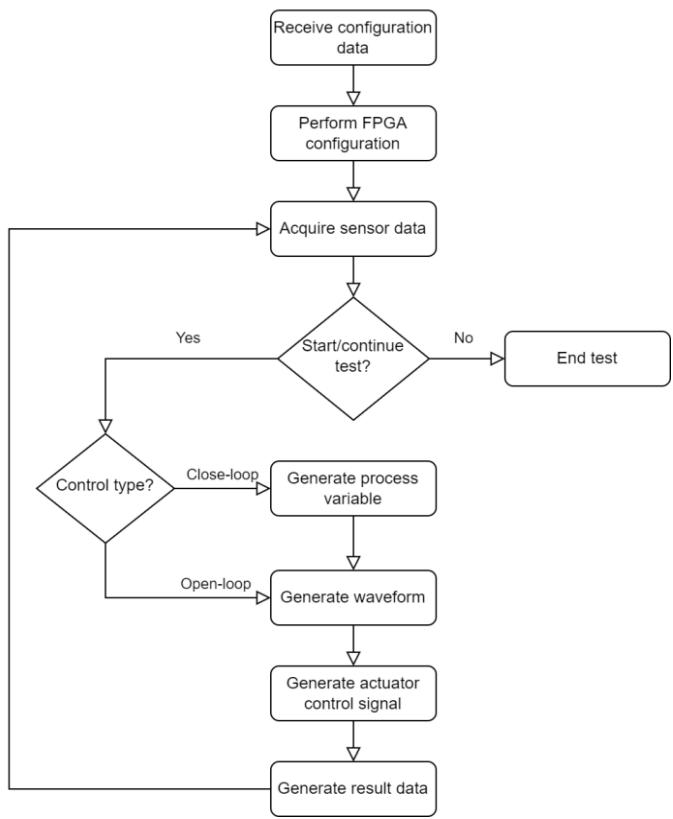

Fig. 2. Material test flow chart

Upon the configuration of Materials Test Machine and Test Controller, the data acquisition process is initiated, thereby causing Test Controller to continuously collect sensor data from Materials Test Machine. The acquired data is sent to the CPU to process, log, and/or communicate to Materials Test Platform. The acquired sensor data may also be used for generating process variables for control of the materials testing in case of close-loop control.

To start the materials testing, CPU sends to FPGA a request to initiate the generation of the control signal for Materials Test Machine. Additionally, FPGA may configure the control loop parameters (e.g., feedback signal based on sensor data that is used for the actuator control) based on the received sensor data.

If the command to start the materials test is received, the process proceeds to the next step, where the controller makes a determination whether the control loop for controlling actuator devices, such as Actuator Device, is an open-loop operation or a closed-loop operation. The determination may be performed based on Test Config Data and/or Machine Config Data or based on user input.

If the FPGA directly controls Materials Test Machine by generating a control signal, such operation is referred to herein as "open-loop" control.

On the other hand, the control signal generation based at least in part on collected sensor data is referred to herein as "closed-loop" control of materials testing. Accordingly, if the process determines that the materials testing is performed in closed-loop control, then the process proceeds to generate process variable based at least in part on the collected sensor data. The process variables are used by FPGA to adjust the actuator devices accordingly.

FPGA generates the input signal to Materials Test Machine on a point-by-point basis for the desired motion. The desired motion may be determined by test config data (e.g., waveform settings, such as frequency and amplitude). Because FPGA has high-speed processing, FPGA may generate a high-precision waveform, where the time period for neighboring points of the control signal may reach several nanoseconds.

In the open-loop operation, each point of the waveform indicates both the desired state and the next state of the materials testing. The digital signal of the point in the waveform is directly provided to DAC (Digital to Analog Converter) or another interface logic (without any control algorithm processing) to generate the corresponding signal output of the waveform for Actuator Device to transition to the desired/next state.

Alternatively, in the closed-loop operation, each point of the waveform indicates the desired state of the materials testing, while the process variable generated based on the sensor data represents the current state of the materials testing. To determine the output for Materials Test Machine, each point of the generated waveform and the process variables are further processed by FPGA's control algorithm, which yields the necessary adjustment output for Actuator Device for the materials testing to transition to the desired state.

In the closed-loop operation, FPGA may be configured to use different control algorithms. For example, test config data or machine config data may indicate which control algorithm for FPGA to use. Based on the config data, FPGA may be configured or reconfigured to use a different control algorithm for the closed-loop operation.

IV. FPGA IMPLEMENTATION OVERVIEW

Described control algorithms, interface (ADC, DAC, Encoder, etc.) drivers are implemented on the FPGA using Verilog hardware description language.

FIG. 3 depicts high level FPGA architecture.

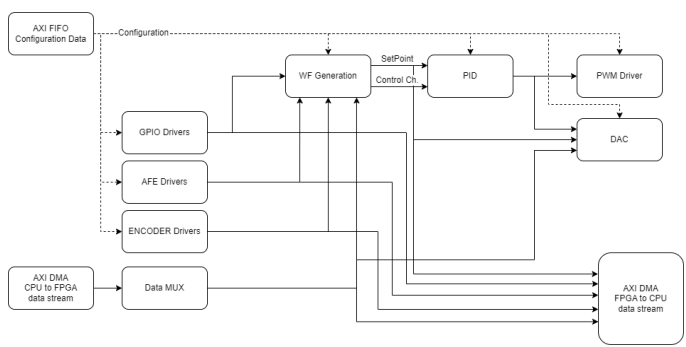

Fig. 3. Material test flow chart

FPGA receives configuration data from the CPU using AXI (Advanced eXtensible Interface) FIFO (First-in, First-

out) interface. The Direct Memory Access (DMA) interface is used to transfer acquired and generated data from the FPGA to CPU. This allows to achieve high speed data transmission.

FPGA implements waveform generation with 64-bit double precision module, achieving high speed and high accuracy profile generation for actuator control.

Depending on the actuator types incorporating into the Material Testing Machine, the FPGA may include different driver, such as pulse-width modulation (PWM) driver to control DC motors, or DAC driver, to control servo-electrical and servo-hydraulic motors, or stepper driver, for corresponding motors.

FPGA implements also control of and data acquisition from general purpose bidirectional digital channels, which are usually used to control different enable mechanisms for the material testing machines, as well as get information from safety devices, such as emergency stop buttons, limit switches, etc.

Inclusion of the data acquisition, profile generation and control algorithms on the same platform, namely, FPGA, allows reliable, high-speed and high accuracy control of different material testing machines.

## V. Results

Material testing machines differ in size, accuracy, speed, and many more parameters, most of which are dependent on the processing and control unit of the machine, i.e. controller.

Most of the vendors comply with 0.5 class of accuracy. But the higher the accuracy the better. Measurement accuracy is dependent on a set of factors, such as ADC resolution, noise immunity, system's internal noise level, etc.

Most of the vendors incorporate 24-bit resolution ADC for force/strain measurement channels, in contrast with the suggested FPGA-based controller, incorporating 32-bit resolution of measurements.

Test results of the first prototypes showed highly competitive numbers for the accuracy of the signal measurements. In comparison with the systems from the main market players, FPGA-based controller has 100x better accuracy (see TABLE 1), plus the highest resolution in the today's market.

TABLE 1.

| Company (UTM/Controller) | Accuracy | Resolution |
|---|---|---|
| MTS (Exceed Series 40 UTM) | 0.5% of reading | 20-bit |
| Instron (5900 Series UTM) | 0.4% of reading | 19-bit |
| Moog (Modular Test Controller) | 0.1% of FSR | 24-bit |
| Suggested Controller | 0.001% of FSR | 32-bit |

Another important parameter is the control loop rate. Current material testing machines incorporate feedback systems for the precise control of the movement, with the most popular control algorithm used as PID.

Control loops are characterized by their response time, i.e. control loop rate. Our research showed that controllers from most of the vendors have 10 kHz control loop rate. With the FPGA-based controller it was possible to achieve 100 kHz control loop rate, decreasing response time down to 10 microsecond.

The suggested controller technology also allows so-called multi-station setup, which means the single controller can control several machines at the same time. This method is quite common in the market and is utilized by several major vendors. The multi-station approach reduces the total cost of the system and is especially beneficial in dynamic/fatigue testing when many samples need to be tested in parallel to reduce the testing time.

The FPGA technology leveraged under the hood allows to control up to 16 single channel stations in parallel, each one providing its own test profile and conducting separate data logging and analysis.